\documentclass[a4paper,11pt]{article}
\pdfoutput=1
\usepackage{jheppub}
\usepackage{amsmath}
\usepackage{amssymb}
\usepackage{color}
\usepackage{xcolor}
\usepackage{graphicx}
\usepackage{hyperref}
\usepackage{textcomp}
\usepackage{xspace,slashed}
\usepackage[utf8]{inputenc}
\usepackage{subcaption}
\usepackage{multirow}
\usepackage{nccmath} 
\usepackage{slashed}

\makeatletter
\renewcommand\@fpheader{}
\renewcommand\@journal{}
\makeatother

\newcommand{\GeV}{\;\mathrm{GeV}}

\newcommand{\ep}{\epsilon}
\newcommand{\oT}{\overline{T}}
\newcommand{\oF}{\overline{\mathcal{F}}}

\newcommand{\OXaff}{Rudolf Peierls Centre for Theoretical Physics, University of Oxford,
  Clarendon Laboratory, Parks Road, Oxford OX1 3PU, U.K.}
\newcommand{\WADaff}{Wadham College, University of Oxford, Oxford OX1 3PN, U.K.}
\newcommand{\UZHaff}{Physik-Institut, Universit\"at Z\"urich, Winterthurerstrasse 190, 8057 Z\"urich, Switzerland}
\newcommand{\FSUaff}{Department of Physics, Florida State University, Tallahassee, FL 32306, USA}

\preprint{
    OUTP-23-10P, ZU-TH 65/23
}

\title{ Two-loop mixed QCD-electroweak amplitudes for $Z+$jet
  production at the LHC: bosonic corrections }
\author[a,c]{Piotr Bargie\l{}a,}
\author[a,b]{Fabrizio Caola,}
\author[a,d]{Herschel Chawdhry,}
\author[a]{Xiao Liu}
\affiliation[a]{\OXaff}
\affiliation[b]{\WADaff}
\affiliation[c]{\UZHaff}
\affiliation[d]{\FSUaff}

\abstract{We present a calculation of the bosonic contribution to the
  two-loop mixed QCD-electroweak scattering amplitudes for $Z$-boson
  production in association with one hard jet at hadron colliders. We
  employ a method to calculate amplitudes in the 't Hooft-Veltman
  scheme that reduces the amount of spurious non-physical information
  needed at intermediate stages of the computation, to keep the
  complexity of the calculation under control. We compute all the
  relevant Feynman integrals numerically using the Auxiliary Mass Flow
  method. We evaluate the two-loop scattering amplitudes on a
  two-dimensional grid in the rapidity and transverse momentum of the
  $Z$ boson, which has been designed to yield a reliable numerical
  sampling of the boosted-$Z$ region. This result provides an
  important building block for improving the theoretical modelling of
  a key background for monojet searches at the LHC.}

\begin{document}

\maketitle
\section{Introduction}
\label{sec:intro}

The production of a $Z$ boson in association with hadronic jets is a
key standard candle at the Large Hadron Collider (LHC). Thanks to its
large production rate and its relatively clean dilepton plus jet
final state, it allows for a multitude of investigations like detector
calibration and luminosity monitoring, validation of Parton Shower
Monte Carlo tools, extraction of fundamental Standard Model (SM)
parameters, as well as high-precision scrutiny of the structure of the
SM and searches for new physics.

Despite the fact that the bulk of the $Z$+jet cross section at the LHC
comes from a region where the jet's transverse momentum is not very
large, $p_{\textrm{T},j}\lesssim 100~{\rm GeV}$, the number of events
where the $Z$ boson is accompanied by a highly-energetic jet is still
quite sizable. This allows for very precise measurements all the way
up to the TeV scale. Indeed, existing experimental
analyses~\cite{ATLAS:2022nrp,CMS:2022ilp} already have a total
uncertainty of just a few percent in the $p_\textrm{T}\sim 200~{\rm
  GeV}$ region and around 10\% in the highly-boosted $p_\textrm{T}\sim
1~{\rm TeV}$ region.  With ever more data being recorded and analysed,
the situation is only going to improve: at the High-Luminosity LHC,
few-percent experimental precision is expected up to transverse
momenta of the order of 1 TeV, and $\mathcal O(10\%)$ precision up to
2.5 TeV~\cite{Lindert:2017olm}.

A good control of $Z$ production in the boosted region allows for
interesting physics explorations. For example, precise data in the
dilepton channel can constrain otherwise elusive dimension-8 Standard
Model Effective Theory (SMEFT) operators~\cite{Boughezal:2022nof},
provided that adequate theoretical predictions are also available.
Also, boosted $Z$-boson production in the $Z$ to invisible channel
provides a key background for monojet searches at the LHC, where one
looks for a high-$p_\textrm{T}$ jet recoiling against missing
energy~\cite{ATLAS:2021kxv,CMS:2021far}. Such a signature is very
interesting because it is quite common in many new physics models,
ranging from weakly coupled dark-matter, to leptoquarks models,
supersymmetric scenarios, or large extra dimensions. 

In the recent past, there has been a large community-wide effort to
improve the theoretical description of boosted vector-boson
production.  In particular, in ref.~\cite{Lindert:2017olm} the authors
provided theoretical predictions that include state-of-the-art NNLO
QCD results~\cite{Gehrmann-DeRidder:2015wbt,
  Gehrmann-DeRidder:2016cdi, Gehrmann-DeRidder:2016jns,
  Boughezal:2015ded, Boughezal:2016isb, Boughezal:2015dva,
  Boughezal:2016dtm, Campbell:2016lzl, Campbell:2017dqk}\footnote{We
note that these references neglect axial-vector contributions for
$Z$+jet production at NNLO. Indeed, the required two-loop scattering
amplitudes for this case were only recently computed in
ref.~\cite{Gehrmann:2022vuk}.}  and NLO electroweak (EWK)
ones~\cite{Denner:2011vu,Denner:2012ts,Kallweit:2015dum,Denner:2009gj}.
In the boosted region, the latter are crucial. Indeed, despite being
suppressed by the weak coupling constant $\alpha$, they are enhanced
by large \emph{Sudakov logarithms} of the form $\frac{\alpha}{4 \pi
  s_w^2} \log^2(\frac{s}{m_V^2})$, where $s_w$ is the sine of the weak
mixing angle, $s$ is a large scale of the process and $m_V$ is the vector
boson mass, see e.g.~\cite{Kuhn:1999de}. At large scales, EWK corrections
then become as large as QCD ones. For this reason, ref.~\cite{Lindert:2017olm}
also included the dominant two-loop electroweak effects coming from Sudakov
logarithms~\cite{Kuhn:2005gv,Kuhn:2005az,Kuhn:2007qc,Kuhn:2007cv}.

Given the size of QCD and EWK corrections, it is also mandatory to
properly control mixed QCD-EWK ones. At present, $\mathcal
O(\alpha_s \alpha)$ corrections to dilepton+jet production are not
known. In ref.~\cite{Lindert:2017olm}, the size of these corrections
was estimated by essentially multiplying NLO QCD and NLO EWK
ones. This prescription is very reasonable at asymptotically-high
scales, since Sudakov logarithms and QCD corrections mostly
factorise. However, at large but finite energies this approximation is
bound to receive corrections. A more rigorous assessment of mixed
QCD-EWK corrections then becomes important. In fact, the lack of exact
$\mathcal O(\alpha_s\alpha)$ corrections is now a major bottleneck
towards highest-precision theoretical predictions in the boosted
region~\cite{Lindert:2017olm}.

Computing $\mathcal O(\alpha_s\alpha)$ corrections to dilepton+jet or
missing-energy+jet production at the LHC poses significant challenges.
First, such processes involve a non-trivial mixed QCD-EWK radiation
pattern, which requires proper regularisation. Achieving this is
complicated if one wants to retain differential information on the
final state. This problem has only recently been solved, and only for
the simplest processes~\cite{Behring:2020cqi, Buccioni:2020cfi,
  Dittmaier:2020vra, Buonocore:2021rxx, Bonciani:2021zzf,
  Buccioni:2022kgy}. Although the techniques used for the
calculation~\cite{Buccioni:2022kgy} could be extended to deal with
more complex processes, this remains a non-trivial task. Second,
mixed QCD-EWK corrections for dilepton+jet or missing-energy+jet production
require non-trivial two-loop scattering amplitudes. These involve both
a complex final state and massive internal virtual particles, which makes
the calculation notoriously difficult.

In this article, we perform a first important step towards the
calculation of mixed QCD-EWK two-loop scattering amplitudes relevant
for boosted dilepton+jet production. To make the problem manageable,
we focus on the production of an on-shell $Z$ boson rather than on the
production of the dilepton final state, with the idea that this is
going to provide the dominant contribution for all observables which
are dominated by the $Z$-pole region (and in particular for the
observables relevant for the boosted region). This allows us to only
consider amplitudes for $2\to2$ scattering rather than the much more
complicated ones relevant for the $2\to 3$ process.  Also, we only
target the boosted region, where vector-boson resonance effects are
not present and hence use of the complex-mass
scheme~\cite{Denner:1999gp, Denner:2005fg, Beneke:2004km,
  Hoang:2004tg, Beneke:1997zp} for EWK corrections becomes less
critical.\footnote{For a discussion of the complex-mass scheme and of
its importance for EWK radiative corrections, see e.g. the recent
review~\cite{Denner:2019vbn} and references therein.}  Finally, as a
first non-trivial step towards the full result here we only consider
bosonic corrections, i.e. we systematically neglect closed
fermion loop corrections.  This allows us to set up a framework for
computing mixed QCD-EWK corrections without the additional
complication of corrections involving top-quark virtual effects, which
cannot a-priori be neglected in the boosted region.\footnote{For the
same reason, we do not-consider $b$-quark induced contributions.} We
believe that our framework could be extended to cover this case as
well, but this warrants an investigation on its own.
Even in this somewhat simplified setup, an analytical calculation of
the amplitude remains challenging. Because of this, we decided to adopt
a semi-numerical approach. Our main result is then an evaluation of
the two-loop amplitudes over a two-dimensional grid that parametrises the
$2\to2$ kinematics. The grid is designed to provide an adequate coverage
of the boosted region. 

The remainder of this paper is organised as follows.  In
sec.~\ref{sec:not} we provide details of our notation and of the
kinematics of the process.  In sec.~\ref{sec:methods} we describe some
methods used in our work i.e. the Lorentz tensor structure of our
scattering amplitude in sec.~\ref{sec:tensors}, and its relation to the
leptonic current at fixed helicity in sec.~\ref{sec:hel}.
In sec.~\ref{sec:comp} we describe our calculation of the bare
amplitudes.  In sec.~\ref{sec:uvir} we discuss the ultraviolet and
infrared structure of our result and we also define one- and two-loop
finite remainders. The latter are the main result of our paper.  In
sec.~\ref{sec:res} we document the checks that we have performed on
our calculation and illustrate our results. Finally, we conclude
in sec.~\ref{sec:concl}.
Our results for the finite remainders are available in
a computer-readable format in the ancillary material that accompanies
this submission.  

\section{Notation and kinematics}
\label{sec:not}
We consider virtual $\mathcal O(\alpha\alpha_s)$ corrections to the
production of the $Z$ boson in association with one hadronic jet.  We
focus on bosonic corrections, i.e. we neglect contributions stemming
from closed fermion loops. We then consider the channels
\begin{equation}
  q + \bar q \to g + Z,~~~~~ q+g \to q + Z,
  \label{eq:genproc}
\end{equation}
where $q$ is either an up- or a down-type (anti) quark. We take the CKM
matrix to be diagonal, and neglect $b$-quark induced contributions.
This way, no virtual top-quark
contributions are present.

All the processes in eq.~\ref{eq:genproc} can be obtained by crossing
the following master amplitude
\begin{equation}
  q(p_1) + \bar{q}(p_2) + g(p_3) + Z(p_4) \to 0\,,
  \label{eq:masterproc}
\end{equation}
where $q$ is either an up or a down quark. In this symmetric notation,
momentum conservation reads
\begin{equation}
	p_1+p_2+p_3+p_4=0\,.
\end{equation}
All external particles are on-shell
\begin{equation}
	p_1^2=p_2^2=p_3^2=0\,, \quad p_4^2=m_Z^2\,.
\end{equation}
The three kinematic Mandelstam invariants of this process 
\begin{equation}
  s_{12} = (p_1+ p_2)^2\,,~ s_{13} = (p_1+p_3)^2\,,~s_{23} = (p_2+p_3)^2 \,,  
\end{equation}
are related by the momentum-conservation relation
\begin{equation}
  s_{12}+s_{13}+s_{23} = m_Z^2\,.
\end{equation}
For physical kinematics, one Mandelstam invariant is positive and two
are negative. Results in the Euclidean region where $m_Z^2<0$ and
$s_{ij}<0$ can be analytically continued to the physical Riemann sheet
by giving a small positive imaginary part to all the invariants, see
ref.~\cite{Gehrmann:2002zr} for a thorough discussion.

For definiteness, we now focus on the $s_{12}>0$, $s_{13},s_{23}<0$ channel.
In the partonic center-of-mass frame, the kinematics can be parametrised
as follows
\begin{equation}
  \begin{split}
    &p_{1} = \frac{\sqrt{s_{12}}}{2}(1,0,0,1)\,,\\
    &p_{2} = \frac{\sqrt{s_{12}}}{2}(1,0,0,-1)\,,\\
    &p_{3,\rm phys} = -p_{3} = p_{t,3}
    \left(\cosh(y_3),\cos\phi,\sin\phi,\sinh(y_3)
    \right)\,\\
    &p_{4,\rm phys} = -p_{4} = 
    \left(m_{t,Z}\cosh(y_Z),-p_{t,Z}\cos\phi,-p_{t,Z}\sin\phi,m_{t,Z}\sinh(y_Z)\right),
  \end{split}
  \label{eq:mompar}
\end{equation}
where $\phi$ is an irrelevant azimuthal angle and $m_{t,Z} =
\sqrt{p_{t,Z}^2+m_Z^2}$. The transverse momenta and rapidities of
the final-state particles are given in terms of Mandelstam invariants
as
\begin{equation}
  p_{t,3} = p_{t,Z} = \sqrt{\frac{s_{13}s_{23}}{s_{12}}}\,,
  ~~~
  y_3 = \frac{1}{2}\ln\left(\frac{s_{23}}{s_{13}}\right),
  ~~~
  y_Z = \frac{1}{2}\ln\left(\frac{s_{12}+s_{23}}{s_{12}+s_{13}}\right),
\end{equation}
with
\begin{equation}
  |y_Z| \le \frac{1}{2}\ln
  \left(
  \frac
      {s_{\rm had}+m_Z^2 + \sqrt{(s_{\rm had}-m_Z^2)^2- 4p_{t,Z}^2 \,s_{\rm had}}}
      {s_{\rm had}+m_Z^2 - \sqrt{(s_{\rm had}-m_Z^2)^2- 4p_{t,Z}^2 \,s_{\rm had}}}
  \right), 
\end{equation}
where $s_{\rm had}$ is the collider center-of-mass energy squared.
In what follows, we will either use $\{s_{13},s_{23}\}$ or $\{p_{t,Z},y_Z\}$
as independent variables. 

To deal with the ultraviolet (UV) and infrared (IR) divergences of the
amplitude, we work in dimensional regularisation and set
$d=4-2\epsilon$.  In particular, we adopt the 't Hooft-Veltman
scheme~\cite{THOOFT1972189}, i.e. we treat all external particles as
purely four-dimensional and the internal ones as
$d=(4-2\ep)$-dimensional. We write the unrenormalised amplitude as
\begin{equation}
  \begin{split}
  \mathcal{A}_b &=
  T^{a_3}_{i_2 i_1}
  \sqrt{\frac{\alpha_{s,b}}{2\pi}} \sqrt{\frac{\alpha_{b}}{2\pi}}
  A_b(d;\{m_{i,b}\},\{s_{ij}\}) \\
  &= T^{a_3}_{i_2 i_1}
  \sqrt{\frac{\alpha_{s,b}}{2\pi}} \sqrt{\frac{\alpha_{b}}{2\pi}} \,
  \epsilon_{3,\mu}(p_3) \epsilon_{4,\nu}(p_4) \, \bar{u}(p_2) A_b^{\mu \nu}(d;\{m_{i,b}\},\{s_{ij}\}) u(p_1) \,,
\end{split}
\label{eq:ampdef}
\end{equation}
where $i_1$, $i_2$, and $a_3$ are colour indices of the quark,
antiquark, and gluon, respectively, $\ep_{3,\mu}$ and $\ep_{4,\nu}$
are the polarization vectors of the massless gluon and of the massive
$Z$ boson, $m_{i,b}$, $i\in\{Z,W\}$ are the (bare) vector boson
masses, and $\alpha_b$, $\alpha_{s,b}$ are the bare electromagnetic
and strong couplings, respectively. The dependence on other
electroweak parameters like the quark isospin or electric charge is
implicitly assumed. Finally, $T^a_{ij}$ is the $SU(3)$ generator in
the fundamental representation, rescaled such that
\begin{equation}
  {\rm Tr}[T^a T^b] = \delta^{a b} .
\end{equation}
We find it convenient to express our results in terms of the quadratic Casimir
invariants $C_A$ and $C_F$, which for $SU(N_c)$ read
\begin{equation}
  C_A = N_c\,,~~~~~ C_F = \frac{N_c^2-1}{2N_c}\,.
  \label{eq:cas}
\end{equation}
In QCD, $C_A=3$ and $C_F = 4/3$. 
The
amplitude in eq.~\ref{eq:ampdef} can be written as a double
perturbative series in the strong and electromagnetic couplings
\begin{equation}
  A_b =
  A^{(0,0)}_{b}
  + \frac{\alpha_{s,b}\,\mu_0^{2\ep}}{2\pi} \, A^{(1,0)}_{b}
  + \frac{\alpha_b\,\mu_0^{2\ep}}{2\pi} \, A^{(0,1)}_{b}
  + \frac{\alpha_{s,b}\,\mu_0^{2\ep}}{2\pi} \frac{\alpha_b\,\mu_0^{2\ep}}{2\pi} \, A^{(1,1)}_{b}
  + \mathcal O(\alpha_{s,b}^2,\alpha_{b}^2)\,,
  \label{eq:ampbare}
\end{equation}
where we have made explicit the dependence on the dimensionful
reference scale $\mu_0$ but have kept the dependence of $A_b$ and
$A_b^{(i,j)}$ on the vector boson masses and on the external
kinematics implicit.  

To obtain the renormalised amplitude $\mathcal A$, we multiply
$\mathcal A_b$ by the external wave-function renormalisation factors
and express the generic bare parameter $g_{i,b}$ in terms of its
renormalised counterpart $g_{i}$.  Schematically,
\begin{equation}
  \mathcal A = \sqrt{Z_q Z_{\bar q} Z_g Z_Z} \times \mathcal A_b\big|_{g_{i,b}\to g_{i}}.
  \label{eq:ren}
\end{equation}
We discuss the renormalisation procedure in more detail in
sec.~\ref{sec:uvir}. Here we only note that we renormalise the strong
coupling in the $\overline{ \rm MS}$ scheme, and all the EWK parameters
in the on-shell scheme. Also, we adopt the $G_\mu$
input parameter scheme, i.e. we choose $\{G_{\mu},m_Z,m_W\}$ as
independent parameters. For our results, we use $G_\mu = 1.16639
\times 10^{-5}$~\cite{ParticleDataGroup:2022pth}.
Similarly to eq.~\ref{eq:ampdef}, we write the renormalised amplitude as
\begin{equation}
  \begin{split}
  \mathcal{A} &=
  T^{a_3}_{i_2 i_1}
  \sqrt{\frac{\alpha_{s}}{2\pi}} \sqrt{\frac{\alpha}{2\pi}}
  A(d;\{m_{i}\},\{s_{ij}\}) \\
  &= T^{a_3}_{i_2 i_1}
  \sqrt{\frac{\alpha_{s}}{2\pi}} \sqrt{\frac{\alpha}{2\pi}} \,
  \epsilon_{3,\mu}(p_3) \epsilon_{4,\nu}(p_4) \, \bar{u}(p_2) A^{\mu \nu}(d;\{m_{i}\},\{s_{ij}\}) u(p_1) \,,
\end{split}
\label{eq:ampdef_ren}
\end{equation}
and expand $A$ as
\begin{equation}
  A =
  A^{(0,0)}
  + \frac{\alpha_{s}}{2\pi} \, A^{(1,0)}
  + \frac{\alpha}{2\pi} \, A^{(0,1)}
  + \frac{\alpha_{s}}{2\pi} \frac{\alpha}{2\pi} \, A^{(1,1)}
  + \mathcal O(\alpha_{s}^2,\alpha^2)\,.
  \label{eq:ArenDefs}
\end{equation}
In these equations, $m_i$ are the renormalised masses, $\alpha$ is the
renormalised electromagnetic coupling and $\alpha_s = \alpha_s(\mu_R)$ is
the renormalised strong coupling with $\mu_R$ the renormalisation scale.
The main result of this article is the computation of $A^{(1,1)}$.

\section{Details of the calculation}
\label{sec:methods}

For further convenience, we provide here some details on the
formalisms used for our calculation.

\subsection{Tensor decomposition}
\label{sec:tensors}

In full generality, the bare scattering amplitudes in eq.~\ref{eq:ampdef}
can be written as
\begin{equation}
  A_b^{(l,m)} = \sum_{k=1}^{n_t}
  \oF^{(l,m)}_{k;b}(d;\{m_{i,b}\},\{s_{ij}\}) \, \times\,
  \oT_k\,,
  \label{eq:ampoT}
\end{equation}
where $\oF^{(l,m)}_{k;b}$ are scalar form factors and $\oT_k =
\bar{v}(p_2) \, \Gamma_k^{\mu \nu} \, u(p_1) \times
\epsilon_{3,\mu}(p_3) \, \epsilon_{4,\nu}(p_4)$, where
$\Gamma_k^{\mu\nu}$ are $n_t$ independent Lorentz tensors. We now
discuss a basis choice for these tensors. First, we note that in this
paper we do not consider corrections coming from closed fermion loops. As a
consequence, there are no anomalous diagrams and we can take
$\gamma_5$ as anticommuting.  Because of this, it is enough to 
study the tensor structure of a vector current, and the axial case will
follow. For a vector current, by simple enumeration one
finds $n_t=39$ independent $\Gamma^{\mu\nu}_i$ structures. Using the
Dirac equation for the external on-shell quarks $\bar{v}_2 \,
\slashed{p}_2 = \slashed{p}_1 u_1 = 0$, as well as the transversality
condition for the massless gluon $\ep_3 \cdot p_3=0$, $n_t$ decreases
by 26. Finally, if one also uses the $Z$-boson
transversality condition $\ep_4\cdot p_4$ and fixes the reference
momentum $q_3$ for the polarisation vector $\ep_3$ to the momentum of
one of the external fermions, one is left with $n_t=7$ independent
$d$-dimensional tensor structures. 

Following ref.~\cite{Gehrmann:2022vuk}, we set $q_3 = p_2$, such that
\begin{equation}
  \ep_3 \cdot p_2=0\,,
  \label{eq:refmom}
\end{equation}
and define the first six tensors as follows\footnote{Note that while our tensors are the same
as the ones in ref.~\cite{Gehrmann:2022vuk}, our numbering differs.}
\begin{align}
  \Gamma_1^{\mu \nu} &= p_1^{\nu} \gamma^{\mu}\,, \;\;\;\;\;
  \Gamma_2^{\mu \nu} = p_1^{\mu} p_1^{\nu} \slashed{p}_3\,, \nonumber \\
  \Gamma_3^{\mu \nu} &= p_2^{\nu} \gamma^{\mu}\,, \;\;\;\;\;
  \Gamma_4^{\mu \nu} = p_1^{\mu} \gamma^{\nu}\,, \nonumber \\
  \Gamma_5^{\mu \nu} &= p_1^{\mu} p_2^{\nu} \slashed{p}_3\,, \;\;
  \Gamma_6^{\mu \nu} = g^{\mu\nu}\slashed{p}_3\,.
  \label{eq:structs}
\end{align}

These six structures span the the whole space of purely
four-dimensional tensors~\cite{Peraro:2019cjj,Peraro:2020sfm}.  We can
then construct the last structure to only have components in the
unphysical $(-2\ep)$-dimensional space. This can be done through a
simple orthogonalisation
procedure~\cite{Peraro:2019cjj,Peraro:2020sfm}, which yields
\begin{equation}
  \oT_7 \,\, = \bar{v}(p_2) \slashed{\epsilon}_3 \slashed{p}_3 \slashed{\epsilon}_4 u(p_1)
  - \frac{1}{s}\left( s(\oT_3+\oT_6-\oT_1) + t\oT_t + u(\oT_1+\oT_4) - 2\oT_5 \right).
  \label{eq:barT}
\end{equation}
Since in strict $d$ dimensions the seven tensors are independent,
there cannot be any cancellation of IR and UV singularities among the
different $\oT_i$.  Hence, the renormalised and IR-regulated form
factor $\oF_{7}^{(i,j)}$ cannot have any poles. Since the
corresponding tensor $\oT_7$ is evanescent by construction, the
contribution of $\oF_{7}^{(i,j)}$ drops from the final result if one
works in the 't Hooft-Veltman scheme and sets $d=4$ after the
renormalisation and IR-regularisation procedure.
Therefore, in the 't Hooft-Veltman scheme, all physical results can be
obtained from the first six tensor structures
alone~\cite{Peraro:2019cjj,Peraro:2020sfm}. We note that the number of
independent tensors coincides with the number of independent helicity
states of the external particles, i.e. it matches the independent
four-dimensional degrees of freedom. A relation between the six tensor
structures in eq.~\ref{eq:structs} and the independent helicity
amplitudes is discussed in the next sub-section.

Since the six tensors in eq.~\eqref{eq:structs} span the physical space,
it is always possible to find coefficients $c_{ik}$ such that
\begin{equation}
  \sum_{\rm pol} \mathcal P_i\,\oT_j \equiv 
  \sum_{\rm pol} \left(\sum_{k=1}^{n_t} c_{ik} \oT_k^\dagger\right) \oT_j = \delta_{ij}.
  \label{eq:proj}
\end{equation}
In fact, solving these equations for $c_{ik}$ is a matter of trivial
algebra. For convenience, we report them in appendix~\ref{app:proj}.
 The
projector operators $\mathcal P_i$ defined in eq.~\ref{eq:proj} can then
be used to extract the $\oF_{k;b}^{(i,j)}$ form factor from the bare
amplitude
\begin{equation}
  \oF^{(i,j)}_{k;b} = \sum_{\rm pol} \mathcal P_k A_b \,.
  \label{eq:projF}
\end{equation}
Finally, we stress that the tensors~\ref{eq:structs}
have been determined under the transversality condition $p_i\cdot \ep_i=0$
and with the reference momentum choice $q_3 = p_2$. It is then important
to use the following expressions for the sum over polarisations of the
external gluon and gauge boson
\begin{equation}
  \begin{split}
    \sum_{\rm pol} \ep_3^{*\mu} \ep_3^\nu &= - g^{\mu\nu}
    + \frac{p_2^\mu p_3^\nu + p_3^\mu p_2^\nu}{p_2 \cdot p_3}\,, \\
    \sum_{\rm pol} \ep_4^{*\mu} \ep_4^\nu &= - g^{\mu\nu}
    + \frac{p_4^\mu p_4^\nu}{m_Z^2}\,.
  \end{split}
  \label{eq:polsum}
\end{equation}

We close this section by discussing how to generalise the above
construction to the case of different left- and right-handed
interactions. As we already mentioned, if we neglect closed fermion
loops there are no anomalous diagrams and one can treat $\gamma_5$ as
anticommuting. This makes such generalisation straightforward. The
amplitude eq.~\ref{eq:ampoT} can be decomposed as
\begin{equation}
  A_b^{(l,m)} = \sum_{c=L,R}\sum_{k=1}^{n_t}
  \oF^{(l,m)}_{k,c;b}(d;\{m_{i,b}\},\{s_{ij}\}) \, \times\,
  \oT_{k,c}\,,
  \label{eq:ampoTLR}
\end{equation}
with
\begin{equation}
  \oT_{i,L/R} = \bar{v}(p_2) \, \Gamma_i^{\mu \nu}
  \left(\frac{1\mp\gamma_5}{2}\right)\, u(p_1) \times
  \epsilon_{3,\mu}(p_3) \, \epsilon_{4,\nu}(p_4),
\end{equation}
and $\Gamma^{\mu\nu}_i$ defined in eq.~\ref{eq:structs}. Projectors
onto the left/right form factors $\oF^{(l,m)}_{i,b,L/R}$ can be
performed using the same procedure explained above, see
eqs~\ref{eq:proj},~\ref{eq:projF}, with the replacements
$\oT^{(\dagger)}_i\to \oT^{(\dagger)}_{i,L/R}$, and
$c_{ik}\to c_{ik}/2$. 
In practice, we always work with vector-current tensors and
projectors, and then dress them with relevant left- and right-handed
couplings. This allows us to effectively halve the various
tensor manipulations on the amplitude.

\subsection{Helicity amplitudes}
\label{sec:hel}
The form factors $\oF_i$ introduced in the previous section can be used
to obtain the helicity amplitudes for the process
\begin{equation}
  q(p_1)+\bar q(p_2) + g(p_3) + \bar l(p_5) + l(p_6) \to 0,
  \label{eq:lept}
\end{equation}
where $l$ and $\bar l$ are a pair of massless leptons.
In the pole approximation~\cite{Stuart:1991xk,Aeppli:1993rs} and
neglecting $\mathcal O(\alpha^2)$ corrections, the amplitude
can be schematically written as
\begin{equation}
  A = i\sum_{\lambda_Z} \frac{\mathcal M^{\rm prod}(\lambda_Z)\times \mathcal M^{\rm dec}(\lambda_Z)}{p_Z^2-m_Z^2 + i m_Z \Gamma_Z} + {\rm non~resonant},
  \label{eq:pole}
\end{equation}
where ``prod''/``dec'' refers to the amplitude for the
production/decay of an on-shell $Z$ boson ($p_Z^2=m_Z^2$) with polarisation
$\lambda_Z$. The numerator in eq.~\ref{eq:pole} can be written in terms
of the tensor structures introduced in sec.~\ref{sec:tensors} as\footnote{The same decomposition holds for both bare and renormalised quantities. We then
omit the ``$b$'' subscript in this section.}
\begin{equation}
  \begin{split}
    \mathcal M \equiv & \sum_{\lambda_Z}
    \mathcal M^{\rm prod}(\lambda_Z)\times \mathcal M^{\rm dec}(\lambda_Z) =
    \\
     &\sum_k 
     \bar{v}(p_2) \Gamma_k^{\alpha\mu}
       \left[\oF_{k,L} \left(\frac{1-\gamma_5}{2}\right)+
      \oF_{k,R}\left(\frac{1+\gamma_5}{2}\right)\right]
       u(p_1) \,\ep_{\alpha}(p_3)
    \times
    \\
    &\left[ -g^{\mu\nu} +
    \frac{p_{Z}^{\mu} p_{Z}^{\nu}}{m_Z^2} \right]
    \times
    \bar{v}(p_5) \gamma_\nu \left[c_{l,L} \left(\frac{1-\gamma_5}{2}\right)+
      c_{l,R}\left(\frac{1+\gamma_5}{2}\right)\right] u(p_6)\,,
  \end{split}
  \label{eq:alep}
\end{equation}
with $p_Z = p_5+p_6 = -p_1-p_2-p_3$, and $p_Z^2=m_Z^2$. In eq.~\ref{eq:alep},
 $c_{l,L}$, $c_{l,R}$
are generalised left- and right-handed couplings that parameterise the
$Zl\bar l$ vertex. At LO, they are given by 
\begin{equation}
  c^{\rm LO}_{l,L} = i e \times \frac{I^3_l - s_w^2 Q_l}{s_w c_w},
  ~~~  c^{\rm LO}_{l,R} = -i e \times \frac{s_w Q_l}{c_w},
\end{equation}
where $I^3_l = \pm 1/2$ is the weak isospin of the lepton $l$, $Q_l$
is its electric charge in units of $e$ ($Q_e=-1$ for an electron), and
$s_w = \sin\theta_w$, $c_w=\cos \theta_w$ with $\theta_w$ the weak
mixing angle. For mixed QCD-EWK corrections, one needs $c_{l,R/L}$ up
to $\mathcal O(\alpha\alpha_s)$. These are well known, see
e.g. refs~\cite{Dittmaier:2009cr,Dittmaier:2015rxo}, so we won't discuss
them further.

The amplitude $\mathcal M$ in eq.~\ref{eq:alep} can be easily projected
to helicity states. To do so, we use the spinor-helicity formalism
(see e.g.~\cite{Dixon:1996wi}) and write left- and right-handed currents
as 
\begin{equation}
  \bar v(p_2) \gamma^\mu \left(\frac{1-\gamma_5}{2}\right) u(p_1) =
  \langle 2\gamma^\mu 1],
    ~~~~
    \bar v(p_2) \gamma^\mu \left(\frac{1+\gamma_5}{2}\right) u(p_1) =
    [ 2\gamma^\mu 1\rangle.
\end{equation}
We also define the helicity of a particle/antiparticle to be equal/opposite
to its chirality, and write the polarisation vector of the \emph{incoming}
gluon as
\begin{equation}
  \ep^\mu_+(p_3) = -\frac{1}{\sqrt2}\frac{[q_3\gamma^\mu 3\rangle}{[q_33]},
    ~~~
    \ep^\mu_-(p_3) = \frac{1}{\sqrt2}\frac{\langle q_3\gamma^\mu3]}
    {\langle q_33\rangle},
\end{equation}
with $q_3$ the reference momentum the gluon, $q_3\cdot \ep_3 =
p_3\cdot \ep_3=0$.  We remind the reader that our tensor $\oT_i$ have
been constructed using the choice $q_3=p_2$. With these assignments,
the helicity amplitudes $\mathcal M_{\vec \lambda}$, $\vec \lambda =
\{\lambda_1,\lambda_2,\lambda_3,\lambda_5,\lambda_6\}$ for the
\emph{all-incoming} process eq.~\ref{eq:lept} read
\begin{equation}
  \begin{split}
    \mathcal{M}_{-+-+-} &= \,\,\,\,\, \frac{c_{l,L}}{\sqrt{2}} \left( \langle 12 \rangle [13]^2 \left( \alpha_{1,L} \langle 536 ] + \alpha_{2,L} \langle 526 ] \right) + \alpha_{3,L} \langle 25 \rangle [13] [36] \right) , \\
      \mathcal{M}_{-+++-} &= \,\,\,\,\, \frac{c_{l,L}}{\sqrt{2}} \left( \langle 23 \rangle^2 [12] \left( \gamma_{1,L} \langle 536 ] + \gamma_{2,L} \langle 516 ] \right) + \gamma_{3,L} \langle 23 \rangle \langle 35 \rangle [16] \right) , \\
	  \mathcal{M}_{+--+-} &= -\frac{c_{l,L}}{\sqrt{2}} \left( [23]^2 \langle 12 \rangle \left( \gamma_{1,R} \langle 536 ] + \gamma_{2,R} \langle 516 ] \right) + \gamma_{3,R} [23] [36] \langle 15 \rangle \right) , \\
	      \mathcal{M}_{+-++-} &= -\frac{c_{l,L}}{\sqrt{2}} \left( [12] \langle 13 \rangle^2 \left( \alpha_{1,R} \langle 536 ] + \alpha_{2,R} \langle 526 ] \right) + \alpha_{3,R} [26] \langle 13 \rangle \langle 35 \rangle \right),
  \end{split}
  \label{eq:hels}
\end{equation}
with 
\begin{equation}
  \begin{split}
    \alpha_{1,i} &= -\oF_{2,i} \,, \qquad
    \alpha_{2,i} = \oF_{5,i} - \oF_{2,i} + \frac{2}{s_{23}} \oF_{4,i} \,, \qquad
    \alpha_{3,i} = 2 \oF_{6,i} - \frac{2 s_{12}}{s_{23}} \oF_{4,i} \,, \\
    \gamma_{1,i} &= \frac{1}{s_{23}} \left( s_{13} \oF_{5,i} + 2 (\oF_{3,i} - \oF_{6,i}) \right) \,, \\
    \gamma_{2,i} &= \frac{1}{s_{23}} \left( s_{13} (\oF_{5,i} - \oF_{2,i}) - 2 (\oF_{1,i} - \oF_{3,i} + \oF_{4,i}) \right) \,, \\
    \gamma_{3,i} &= -\frac{2}{s_{23}} \left( s_{23} \oF_{6,i} + s_{12} \oF_{4,i} \right) \,,
  \end{split}
\end{equation}
and $i=L,R$. The amplitudes with $\{\lambda_5,\lambda_6\} = \{-,+\}$
can obtained from the expressions \ref{eq:hels} for
$\{\lambda_5,\lambda_6\} = \{-,+\}$ through the replacements
$|5\rangle\to |5]$, $|6]\to |6\rangle$, $c_{l,L}\to c_{l,R}$. Using
    these results, it is straightforward to reconstruct any amplitude
    from the $\oF$ form factors. Because of this, we will focus on the
    latter in what follows.
We conclude this section by pointing out that eq.~\ref{eq:hels} is
in agreement with eqns (5.6-5.17) of ref.~\cite{Gehrmann:2022vuk} once
all the small differences in notation are accounted for.

\subsection{Computation of the bare amplitude}
\label{sec:comp}
\begin{figure}[h]
  \centering
  \includegraphics[width=0.9\textwidth]{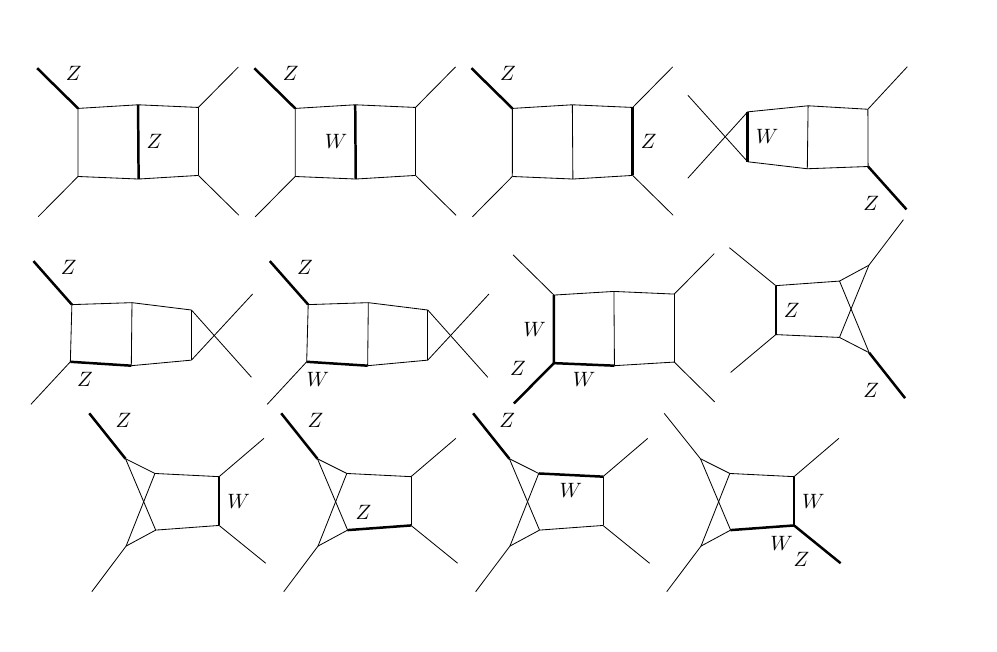}
  \caption{Example Feynman diagrams corresponding to top integral topologies.}
  \label{fig:diags}
\end{figure}

\begin{table}[h]
  \centering
  \begin{tabular}{ p{4.5cm}||p{1.0cm}|p{1.0cm}|p{1.0cm}|p{1.0cm} }
    & ${A}_b^{(0,0)}$ & ${A}_b^{(1,0)}$ & ${A}_b^{(0,1)}$ & ${A}_b^{(1,1)}$ \\
    \hline
    \hline
    \# non-vanishing diagrams & 2 & 11 & 27 & 462 \\
    \hline
    \# integral topologies & 0 & 1 & 4 & 18 \\
    \hline
    \# scalar integrals & 0 & 105 & 275 & 60968 \\
  \end{tabular}
  \caption{Complexity comparison at different loop orders.}
  \label{tab:comp}
\end{table}

We now describe the calculation of the bare two-loop mixed QCD-EWK
factors $\oF$.  We start with generating, using
\texttt{Qgraf}~\cite{Nogueira:1991ex}, all the relevant Feynman
diagrams contributing to the bare amplitude $A_b^{(1,1)}$, see
fig.~\ref{fig:diags} for few representatives.  As we already
mentioned, we neglect closed fermion loops and bottom-induced
contributions.  We are left with 462 non-vanishing Feynman diagrams
$(\text{FD})^{(1,1)}_f$,
\begin{equation}
  {A}_b^{(1,1)} = \sum_{f=1}^{462} (\text{FD})^{(1,1)}_f \,.
\end{equation}
Comparing to lower orders, (see tab.~\ref{tab:comp}) the complexity
grows significantly, thus making the efficiency of the calculation
crucial.  For this reason, we first perform a useful
parallelisation criterion, and only later the projection of the
amplitude onto the tensor structures introduced in sec.~\ref{sec:tensors}.
To this end, we split the Feynman diagrams by their graph
structure.

In the two-loop four-point amplitude, each Feynman diagram can have at
most 7 virtual propagators.  We will refer to diagrams containing a
set of 7 different virtual propagators as \textit{top sector}
diagrams.  Given the kinematics of the problem, there are 9
independent Lorentz invariants involving loop momenta $k_i$, $i=1,2$.
Out of these, 3 are of the type $k_i \cdot k_j$ and 6 of the type $p_i
\cdot k_j$.  These 9 invariants can be linearly related to the 7 top
sector propagators complemented with 2 additional irreducible scalar
products (ISPs), which for convenience we also choose to be of the
propagator type.  We will refer to this set of 9 propagators as an
\textit{integral topology}.  One can find a minimal set of integral
topologies onto which all Feynman diagrams can be mapped.  For our
case, we require 18 basic topologies, as well as their 61
crossings. We report the definition of the 18 basic topologies in
appendix~\ref{app:topo}, and in computer-readable format in an
ancillary file provided with this submission.
Feynman diagrams with a smaller number of virtual propagators can be
obtained by pinching top sector diagrams, which often makes their
mapping onto an integral topology not unique.  In practice, we perform
this mapping by finding an appropriate loop momenta shift with
\texttt{Reduze 2}~\cite{vonManteuffel:2012np,Studerus:2009ye}. As a result, we
split the whole set of Feynman diagrams contributing to
$\mathcal{A}_b^{(1,1)}$ by integral topology type $t$
\begin{equation}
  {A}_b^{(1,1)} = \sum_{t \in \text{topo}} {A}^{(1,1)}_{b,t} \,.
\end{equation}
Common structures of different integral topologies are typically
revealed only after loop-momenta integration and/or expansion in
$\ep$. Hence, grouping Feynman diagrams by topology and performing
preliminary manipulations separately for each topology provides
an efficient parallelisation strategy.

To actually compute the amplitude, we work in the Feynman gauge. After
substituting the Feynman rules, we immediately perform the colour algebra
and write our result in terms of $C_A$ and $C_F$, see eq.~\ref{eq:cas}.
Next, we project on the 12 form factors
$\oF_{1,...,6,L/R;b}$ defined in sec.~\ref{sec:tensors}.
To do so, we apply the $\mathcal{P}_i$ projectors (see
sec.~\ref{sec:tensors}) and evaluate all required traces of Dirac
gamma matrices in $d$-dimensions with
\texttt{Form}~\cite{Vermaseren:2000nd}, treating $\gamma_5$ as anticommuting.
At the end, each form factor can be written as
\begin{equation}
  \oF^{(1,1)}_{i,L/R;b} = \sum_t \sum_{f \, \in \, \text{FD}_t} \int
  \frac{d^dk_1}{(2\pi)^d} \frac{d^dk_2}{(2\pi)^d} \, \frac{\mathcal
    N_{i,L/R,f}(d;\{m_i\},\{p_i\cdot p_j\},\{p_i\cdot k_j\},\{k_i \cdot
    k_j\})}{\mathcal D_{t,1}^{n_{f,1}} \dots \mathcal
    D_{t,7}^{n_{f,7}}} \,,
  \label{eq:f_from_dia}
\end{equation}
where the first sum runs over all the relevant topologies, FD$_t$ is
the set of Feynman diagrams mapped onto a given topology $t$,
$\mathcal N$ is a polynomial in the space-time dimension $d$, the $Z$
and $W$ masses and scalar products involving both loop and external
momenta, $\mathcal D_{t,i}$ is the $i$-th denominator factor of the
topology $t$ (see appendix~\ref{app:topo} for their explicit definition),
and $n_{f,i} \in \{0,1,2\}$.
For convenience, we linearly relate all the 9 independent kinematic
invariants involving loop momenta, which are present in numerators
$\mathcal N$, to the complete set of propagators
$\vec{\mathcal D}_{t}$, fixed by the integral topology $t$, such that
eq.~\ref{eq:f_from_dia} assumes the form
\begin{equation}
  \oF^{(1,1)}_{i,L/R;b} = \sum_t\sum_{\vec{n} \, \in \, \text{int
      set}} \widetilde{\mathcal N}_{i,t,\vec{n}}(d;m_i,\{p_i \cdot p_j\}) \,
  \mathcal{I}_{t,\vec{n}} \,, \qquad n_i \in [-4,2] \cap \mathbb{Z}
  \,,
\end{equation}
with now
\begin{equation}
	\mathcal{I}_{t,\vec{n}} \equiv
	\int \frac{d^dk_1}{(2\pi)^d} \frac{d^dk_2}{(2\pi)^d} \,
	\frac{1}{\mathcal{D}_{t,1}^{n_1} \cdots \mathcal{D}_{t,9}^{n_9}} \,.
\end{equation}
It is well-known that the $\mathcal I_{t,\vec n}$ integrals are not
independent. Indeed, they satisfy so-called Integration-By-Parts (IBP)
identities~\cite{Chetyrkin:1981qh}. Typically, only after the form
factors are expressed in terms of a minimal set of independent
\emph{master integrals} (MIs) one sees a significant reduction in the
complexity of the $\widetilde{\mathcal N}$ coefficients, as redundancies
tend to be minimised.

Since in our problem there are many mass scales, satisfactorily
performing a fully-analytic reduction of the amplitude in terms of MIs
is complicated. In particular, the step of expressing the various
integrals in terms of MIs typically introduces complex rational
functions.  Although these are expected to greatly simplify when all
the pieces of the amplitude are combined together, achieving such a
simplification is non-trivial. Fortunately, we can simplify our
calculation by setting the $W$ and $Z$ masses to numerical values.
For numerical efficiency, we choose the $m_W^2/m_Z^2$ ratio to be a
simple rational number, while being close enough to its actual value.
In particular, we choose
\begin{equation}
  \begin{split}
    m_Z   &=  91.1876 \GeV \,, \\    
    m_W   &= \sqrt{\frac{7}{9}} m_Z = 80.4199 \GeV \,.
  \end{split}
\end{equation}
Such a choice for the $W$ mass differ from its actual value by few
permille, and it is fully adequate for our phenomenological purposes.
Setting the $Z$ and $W$ mass to numbers vastly reduces the parametric
complexity of the integral structure. Keeping both the
dimensional-regularisation parameter $\ep$ and the kinematics
invariant $\{s_{13},s_{23}\}$ symbolic, we generate IBP identities by
\texttt{LiteRed}~\cite{Lee:2012cn, Lee:2013mka} and solve them by the
finite field arithmetic~\cite{vonManteuffel:2014ixa, Peraro:2016wsq}
implemented in \texttt{FiniteFlow}~\cite{Peraro:2019svx}. We
also employ the method described in ref.~\cite{Liu:2023cgs} to exploit
linear relations among the reduction coefficients, which can
effectively utilize their common structures to reduce the number of
finite-field samples. Comparing to the traditional reconstruction
strategy, this method can improve the computational efficiency by
approximately an order of magnitude in our computation. 

After reducing the amplitude to MIs, we evaluate the latter using the
power-expansion-based differential equations
method~\cite{Kotikov:1990kg, Czakon:2008zk}. To this end, we construct
a differential equation for master integrals with respect to the
kinematics invariants $s_{13}$ and $s_{23}$. Schematically,
\begin{align}
  \partial_x
  \vec{I} &= \mathbf{A}_x(\epsilon, s_{13}, s_{23})\vec I,\quad\quad x \in\{s_{13},s_{23}\}
\end{align}
where $\vec I$ is a vector containing all the MIs and ${\mathbf A}_x$,
are matrices. Such a writing is possible since the MIs provide a basis
not only for integrals but also for all their possible derivatives.
To solve the differential equations, we first compute the boundary
conditions by numerically evaluating the MIs at the following regular
point in $s_{12}$ scattering region:
\begin{align}
  \frac{s_{13}}{m_Z^2}=-\frac{31}{11}, \quad \frac{s_{23}}{m_Z^2} = -\frac{13}{7}.
\end{align}
To do this, we use \texttt{AMFlow}~\cite{Liu:2022chg} which implements
the auxiliary mass flow
method~\cite{Liu:2017jxz,Bronnum-Hansen:2020mzk,Liu:2021wks}, which
has already proven very successful in many complex calculations, see
e.g.~\cite{Chen:2022vzo,Chen:2022mre,FebresCordero:2023gjh}.  With
differential equations and boundary conditions in hand, the master
integrals have already been fully determined. What we need to do in
practice is just to use the differential equations solver in
\texttt{AMFlow} to solve them from the boundary point to any desired
phase space points.  We compute the $\ep$ expansion of our results by
fitting 10 numerical samples obtained using specific numerical values
for $\ep$~\cite{Liu:2022chg}.  To make sure that we retain enough
numerical precision even in case of large cancellations among terms,
at this stage we evaluate all the MIs with about 120-digit precision.
From these 10 numerical samples, we reconstruct the $\ep$ expansion of
the result directly at the level of the form factors, without first
reconstructing the analogous expansion for the MIs. After this procedure,
we expect our leading pole to have very high precision of about 50 digits
or better. We then lose a bit of precision for each
subsequent order in $\ep$, but still we estimate that our final part
retains about 20-digit precision or better. 

In this way, we are able to numerically evaluate all the $\oF$ form
factors at a given phase-space point in a robust and relatively
efficient way. Contrary to a fully-analytic result, such an evaluation
cannot be kinematically crossed. Therefore, we need in principle to
provide 6 separate results for the independent channels $u \, \bar{u}
\to g \, Z$, $u \, g \to u \, Z$, $g \, u \to u \, Z$, $d \, \bar{d}
\to g \, Z$, $d \, g \to d \, Z$, $g \, d \to d \, Z$. In practice, we
evaluate our result on a two-dimensional $\{p_{t,Z},y_Z\}$ grid that
is symmetric under $p_{1}\leftrightarrow p_{2}$ exchange,
see sec.~\ref{sec:res}. As a consequence,
we only need to compute results for the four channels $u \, \bar{u}
\to g \, Z$, $u \, g \to u \, Z$, $d \, \bar{d}
\to g \, Z$, $d \, g \to d \, Z$.

\section{UV renormalisation and IR regularisation}
\label{sec:uvir}

The bare two-loop mixed QCD-EWK amplitude ${A}^{(1,1)}_{b}$ has pole
singularities in $\ep$ starting at $\ep^{-4}$, stemming from UV and IR
divergences. As we mentioned in sec.~\ref{sec:not}, we renormalise the
strong coupling in the $\overline{ \rm MS}$ scheme, all the EWK
parameters in the on-shell scheme, and we adopt the $G_\mu$
input-parameter scheme. To renormalise our result, one has to write
the bare parameters in terms of the renormalised ones and to multiply
by the relevant $Z$-factors, see eq.~\ref{eq:ren}. To be more concrete,
we expand our bare form factors as
\begin{equation}
  \begin{split}
  \oF_{i,c;b} =
  \sqrt{\frac{\alpha_{s,b}}{2\pi}} \sqrt{\frac{\alpha_{b}}{2\pi}}
  \bigg[
  \oF^{(0,0)}_{i,c;b}
  + \frac{\alpha_{s,b}\,\mu_0^{2\ep}}{2\pi} \, \oF^{(1,0)}_{i,c;b}
  + \frac{\alpha_b\,\mu_0^{2\ep}}{2\pi} \, \oF^{(0,1)}_{i,c;b}
  +
  \\
  \frac{\alpha_{s,b}\,\mu_0^{2\ep}}{2\pi} \frac{\alpha_b\,\mu_0^{2\ep}}{2\pi} \, \oF^{(1,1)}_{i,c;b}
  + \mathcal O(\alpha_{s,b}^2,\alpha_{b}^2)\bigg]\,,
  \end{split}
\end{equation}
with $i=1,...,6$ and $c=L,R$, see eq.~\ref{eq:ampbare}. We also define
renormalised form factors through the expansion
\begin{equation}
  \begin{split}
  \oF_{i,c} =
  \sqrt{\frac{\alpha_{s}}{2\pi}} \sqrt{\frac{\alpha}{2\pi}}
  \bigg[
  \oF^{(0,0)}_{i,c}
  + \frac{\alpha_{s}}{2\pi} \, \oF^{(1,0)}_{i,c}
  + \frac{\alpha}{2\pi} \, \oF^{(0,1)}_{i,c}
  +
  \frac{\alpha_{s}}{2\pi} \frac{\alpha}{2\pi} \, \oF^{(1,1)}_{i,c}
  + \mathcal O(\alpha_{s}^2,\alpha^2)\bigg]\,,
  \end{split}
\end{equation}
with now $\alpha_s=\alpha_s(\mu)$. Similarly to eq.~\ref{eq:ren}, bare
and renormalised form factors are schematically related by
\begin{equation}
  \oF_{i,c} = \sqrt{Z_{q,c} Z_{\bar q,c} Z_g Z_Z} \times \mathcal \oF_{b,c}\big|_{g_{i,b}\to g_{i}}.
  \label{eq:renF}
\end{equation}

At $\mathcal
O(\alpha_s)$, all the $Z_i$ factors are equal to one and the QCD
renormalisation procedure amounts to writing the bare coupling
$\alpha_{s,b}$ in terms of the $\overline{\rm MS}$ one as
\begin{equation}
  S_{\epsilon}\mu_0^{2\ep}\alpha_{s,b} = \mu^{2\ep} \alpha_s
  \left[1-\frac{\beta_0}{\ep}\left(\frac{\alpha_s}{2\pi}\right)+
    \mathcal O(\alpha_s^2)\right],
  \label{eq:asren}
\end{equation}
with
$S_{\epsilon} = (4\pi)^{\epsilon}e^{-\gamma_E \epsilon}$ and $\alpha_s =
\alpha_s(\mu)$. If we neglect contributions coming from closed fermion loops,
$\beta_0$ reads
\begin{equation}
  \beta_0 =\frac{11}{6}C_A.
\end{equation}
The $\mathcal O(\alpha)$ EWK renormalisation procedure is more
complicated. However, since at this order neither the strong coupling
nor $Z_g$ receive corrections, it is formally identical to the
renormalisation of the $q\bar q Z$ vertex. To this order, we can write
\begin{equation}
  \oF_{i,c} = \left[1+\frac{\alpha}{2\pi}\delta_{{\rm
      UV},c}^{(0,1)}\right]\oF_{i,c;b},
\end{equation}
with~\cite{Dittmaier:2009cr}\footnote{We note that $\{L,R\}$ here
correspond to $\{-,+\}$ in ref.~\cite{Dittmaier:2009cr}.
Also, we note that the UV counterterms depend on the numerical value
of the Higgs mass, that we set to $m_H=125 \GeV$.}
\begin{equation}
  \delta_{{\rm UV},c}^{(0,1)} = \frac{1}{2}\left(2\delta_{Z_{q,c}} +
  \delta_{Z_{ZZ}}- \frac{Q_q}{g_{q,c}}\delta_{Z_{\gamma Z}}\right)+
  \frac{\delta_{g_{q,c}}}{g_{q,c}}.
  \label{eq:deltaUV}
\end{equation}
In this equation,
\begin{equation}
  g_{q,L} = \frac{I^3_{q}-s_w^2 Q_{q}}{s_wc_w} \,,\quad\quad
  g_{q,R} = -\frac{s_wQ_{q}}{c_w} \,,
  \label{eq:gcoup}
\end{equation}
where $Q$ is the electric charge of the external quark in units of $e$
($Q_{up} = \frac{2}{3}$, $Q_{dn} =-\frac{1}{3}$), $I^3_q$ is the third
component of its weak isospin ($I^3_{up}=\frac{1}{2}$,$I^3_{dn} = -
\frac{1}{2}$), $c_w = \frac{m_W}{m_Z}$ and $s_w=\sqrt{1-c_w^2}$.  Note
the $\gamma Z$ kinetic mixing term $\delta_{Z_{\gamma Z}}$ in
eq.~\ref{eq:deltaUV}, which can be accounted for by a simple
refactoring of the overall coupling.  Note also that, contrarily to
the $\overline{ \rm MS}$ renormalization scheme, in the on-shell
scheme, the counterterms have a non-trivial $\ep$ expansion. Their
explicit expression up to $\mathcal O(\ep^0)$ can be found e.g. in
ref.~\cite{Denner:1991kt}.

At $\mathcal O(\alpha\alpha_s)$, there is a non-trivial interplay
between QCD and EWK renormalisation. However, the situation
drastically simplifies if one neglects contributions coming from
closed fermion loops. Indeed, in this case still $Z_g=1$. Also, the
QCD and EWK renormalisation procedures almost decouple from each
other.  Precisely, only the quark wave-function renormalisation factor
$Z_q$ receives $\mathcal O(\alpha\alpha_s)$ contributions. For left-handed
quarks, it reads~\cite{Buccioni:2020cfi,Behring:2020cqi}
\begin{equation}
  \begin{split}
    Z_{q,L} ={}& 1 -\left(\frac{\alpha}{2 \pi}
    \frac{(4\pi)^{\ep}}{\mu^{\ep}}\right) \frac{1 - \ep}{(2 - \ep) \ep}
    \Gamma(1 + \ep) \left(g_{q,L}^2
    \left(\frac{m_Z^2}{\mu^2}\right)^{-\ep} +\frac{1}{2 s_w^2}
    \left(\frac{m_W^2}{\mu^2}\right)^{-\ep} \right) \\ &
    +\left(\frac{\alpha}{2 \pi} \frac{\alpha_{s}(\mu)}{2 \pi}
    \frac{(4\pi)^{2\ep}}{\mu^{2\ep}} \right) C_F \frac{(3 - 2 \ep) (1 -
      3 \ep)}{4 \ep (2 - \ep) (1 - 2 \ep)} \Gamma(1 - \ep) \Gamma(1 +
    \ep) \Gamma(1 + 2 \ep) \\ & \hphantom{+~} \times \left(g_{q,L}^2
    \left(\frac{m_Z^2}{\mu^2}\right)^{-2 \ep} +\frac{1}{2 s_w^2}
    \left(\frac{m_W^2}{\mu^2}\right)^{-2 \ep} \right) + \mathcal
    O(\alpha_s^2,\alpha^2) =
    \\
    & 1 + \frac{\alpha}{2\pi}
    \delta^{(0,1)}_{Z_{q,L}} +\frac{\alpha}{2\pi}\frac{\alpha_s(\mu)}{2\pi}
    \delta^{(1,1)}_{Z_{q,L}}+   O(\alpha_s^2,\alpha^2)\,.
  \end{split}
  \label{eq:ZqL}
\end{equation}
The analogous result for right-handed quarks can be obtained from
eq.~\ref{eq:ZqL} by substituting $g_{q,L}\to g_{g,R}$ and removing the
$W$ contribution. 

Combining everything together, we can then write the renormalised form
factors in terms of their bare counterparts as 
\begin{equation}
  \begin{split}
    \oF^{(0,0)}_{i,c} &= S_\ep^{-1/2} \oF^{(0,0)}_{i,c;b} \,,
    \\ \oF^{(1,0)}_{i,c} &= S_\ep^{-1/2} \left( S_\ep^{-1}
    \oF^{(1,0)}_{i,c;b} - \frac{\beta_0}{2\ep} \,\oF^{(0,0)}_{i,c;b}
    \right) \,, \\ \oF^{(0,1)}_{i,c} &= S_\ep^{-1/2} \left(
    \oF^{(0,1)}_{i,c;b} + \delta_{{\rm UV},c}^{(0,1)}\,
    \oF^{(0,0)}_{i,c;b} \right) \,, \\ \oF^{(1,1)}_{i,c} &=
    S_\ep^{-1/2} \left( S_\ep^{-1} \oF^{(1,1)}_{i,c;b} -
    \frac{\beta_0}{2\ep} \oF^{(0,1)}_{i,c;b} + \delta_{{\rm
        UV},c}^{(0,1)}\, S_\ep^{-1} \oF^{(1,0)}_{i,c;b} \right) \\ &+
    S_\ep^{-1/2} \left( \delta_{Z_{q,c}}^{(1,1)} - \frac{\beta_0}{2\ep}
    \delta_{{\rm UV},c}^{(0,1)} \right) \oF^{(0,0)}_{i,c;b} \,.
  \end{split}
  \label{eq:expl_renorm}
\end{equation}
We note that since $\oF_{i,c;b}^{(1,0)}$ contains poles, in principle
we require $\delta_{{\rm UV},c}^{(0,1)}$ at higher orders in
$\ep$. In practice, this is not the case since they always decouple
from physical quantities. To see how this comes about, we first need 
to discuss the structure of IR divergences. 

The soft and collinear structure of UV-renormalised one-loop amplitudes
can be immediately extracted from Catani's formula~\cite{Catani:1998bh}.
In our case, we write
\begin{equation}
  \begin{split}
    \oF_{i,c}^{(1,0)} = \mathcal I^{(1,0)}\, \oF_{i,c}^{(0,0)} +
    \oF_{i,c}^{(1,0),\rm fin}, \quad \quad \quad 
    \oF_{i,c}^{(0,1)} = \mathcal I_q^{(0,1)}\, \oF_{i,c}^{(0,0)} +
    \oF_{i,c}^{(0,1),\rm fin}, 
  \end{split}
  \label{eq:cat1L}
\end{equation}
where $\oF_{i,c}^{(i,j),\rm fin}$ are finite remainders and
\begin{equation}
  \begin{split}
    \mathcal I^{(1,0)} &=
    \frac{e^{\gamma_E\ep}}{2\Gamma(1-\ep)} \Bigg\{
    \left(\frac{\mu^2}{-s_{q\bar q}-i\varepsilon}\right)^\epsilon
    (C_A-2C_F)\left[\frac{1}{\epsilon^2}+\frac{3}{2\epsilon}\right]
    \\ &- \left[\left(\frac{\mu^2}{-s_{q g}-i\varepsilon}\right)^\epsilon +
      \left(\frac{\mu^2}{-s_{\bar q g}-i\varepsilon}\right)^\epsilon\right]
    \left[C_A\left(\frac{1}{\epsilon^2}+\frac{3}{4\epsilon}\right) +
      \frac{\beta_0}{2\epsilon}\right]\Bigg\} \,, \\ \mathcal
    I^{(0,1)}_q &= -\frac{S_\ep
      e^{\gamma_E\ep}}{\Gamma(1-\ep)}
    \left(\frac{\mu^2}{-s_{q\bar q}-i\varepsilon}\right)^\epsilon Q_{q}^2
    \left[\frac{1}{\epsilon^2}+\frac{3}{2\epsilon}\right] .
    \label{eq:poles1L}
  \end{split}
\end{equation}
In these equations, $s_{ij} = (p_i + p_j)^2$, for all-incoming
kinematics eq.~\ref{eq:masterproc}. At mixed QCD-EWK order, we follow
refs~\cite{Behring:2020cqi,Buccioni:2022kgy} and write
\begin{equation}
  \oF_{i,c}^{(1,1)} = \mathcal I_q^{(1,1)} \oF_{i,c}^{(0,0)}+
  \mathcal I^{(1,0)}\oF_{i,c}^{(0,1),\rm fin} + 
  \mathcal I_q^{(0,1)}\oF_{i,c}^{(1,0),\rm fin} +
  \oF^{(1,1),\rm fin}_{i,c},
  \label{eq:cat2L}
\end{equation}
where again $\oF^{(1,1),\rm fin}_{i,c}$ is finite and\footnote{We
note that our definition of finite remainders is slightly different
from the one in refs~\cite{Behring:2020cqi,Buccioni:2022kgy}.}
\begin{equation}
  \mathcal I_q^{(1,1)} = \mathcal I^{(1,0)} \mathcal I_q^{(0,1)}+
  \frac{S_\ep e^{\gamma_E\epsilon}}{\Gamma(1-\epsilon)}  
  \frac{1}{\ep}
  C_F Q_{q}^2
  \left(\frac{\pi^2}{2} - 6 \zeta_3 - \frac{3}{8}\right).
\end{equation}
The finite remainders $\oF_{i,c}^{(1,1),\rm fin}$ are the main results
of this work. We report them on a numerical grid in $\{p_{t,Z},y_{Z}\}$
in the ancillary files. For convenience, we also include the lower order
results $\oF_{i,c}^{(0,0),\rm fin} = \oF_{i,c}^{(0,0)}$,
$\oF_{i,c}^{(0,1),\rm fin}$, and $\oF_{i,c}^{(1,0),\rm fin}$. 

We conclude this section by explicitly illustrate how higher orders in
$\ep$ in the one-loop UV EWK renormalisation counterterm $\delta_{{\rm
    UV},c}^{(0,1)}$ decouple from our finite remainder
$\oF_{i,c}^{(1,1),\rm fin}$. Substituting eqs~\ref{eq:expl_renorm} in
the finite-remainder definitions eqs~\ref{eq:cat1L},~\ref{eq:cat2L},
it is straightforward to find that the $\delta_{{\rm UV},c}^{(0,1)}$
contribution to $\oF_{i,c}^{(1,1),\rm fin}$ reads
\begin{equation}
  \begin{split}
    \oF_{i,c}^{(1,1),\rm fin} = &\delta_{{\rm UV},c}^{(0,1)}S_\ep^{-1/2}
    \left[
      S_\ep^{-1}\oF_{i,c;b}^{(1,0)}-\frac{\beta_0}{2\ep} \oF_{i,c;b}^{(0,0)}
      -\mathcal I^{(1,0)} \oF_{i,c;b}^{(0,0)}
      \right] + ... = \\
    & \delta_{{\rm UV},c}^{(0,1)} \, \oF_{i,c}^{(1,0),\rm fin} + ...\,,
    \end{split}
\end{equation}
where ellipses stand for terms that do not contain $\delta_{{\rm
    UV},c}^{(0,1)}$. Since $\oF_{i,c}^{(1,0),\rm fin}$ is finite,
higher orders in the $\ep$ expansion of $\delta_{{\rm UV},c}^{(0,1)}$
decouple from the finite remainder $\oF_{i,c}^{(1,1),\rm fin}$ for
$\ep = 0$.

\section{Checks and final results}
\label{sec:res}
As we have mentioned in sec.~\ref{sec:comp}, we decided to compute the
mixed QCD-EWK amplitudes numerically. In order for our result to be
useful, we have performed the numerical evaluation on a dense-enough grid
in $\{p_{t,Z},y_{Z}\}$ (see sec.~\ref{sec:not}). We focus on the boosted-$Z$
region, where there are no thresholds. Because of this, we choose our grid
to be logarithmically uniform in $p_{t,Z}$ and linearly uniform in
$y_Z$, with dimension 40x41. Specifically, 
\begin{equation}
  \begin{split}
    p_{T,Z,n} &= 200 \cdot 10^{n/39} \,, \qquad n \in [0,39] \cap \mathbb{Z} \,, \\
    y_{Z,m} &= \frac{m}{4}-5 \,, \qquad\qquad m \in [0,40] \cap \mathbb{Z} \,.
  \end{split}
  \label{eq:grid}
\end{equation}
The kinematic ranges are chosen to allow for studies both at the LHC
and at future colliders. The grid~\ref{eq:grid} is
symmetric under $y_Z\to - y_Z$, which allows us to only
consider a subset of the relevant partonic channels, see the discussion at
the end of sec.~\ref{sec:comp}. Note that both the grid parametrization
as well as the change of variables to Mandelstam invariants
$\{p_{t,Z},y_Z\}\to\{s_{13},s_{23}\}$ are not rational.
In order to control our numerical precision, we
evaluate all the required Feynman integrals at the resulting numerical
$\{s_{13},s_{23}\}$ grid rationalized within 8-digit agreement. The
rationalised version of the grid~\ref{eq:grid} can be found in an
ancillary file. 
To check the suitability of our grid for
phenomenological studies, we have compared LO predictions for $a)$ the
total cross section for $Z+j$ production with $p_{t,Z}>200 \GeV$, $b)$
the differential distributions ${\rm d}\sigma/{\rm d}p_{t,Z}$, ${\rm
  d}\sigma/{\rm d}y_Z$, and $c)$ the double differential distribution
${\rm d}\sigma/{\rm d} p_{t,Z}/{\rm d}y_Z$ obtained from our grids
against \texttt{MCFM 6.8}~\cite{Campbell:2015qma} and found
satisfactory agreement.

Before presenting our final results, we describe various checks of our
calculation that we have performed. First, as a byproduct of our
computation we have re-computed the tree-level, one-loop QCD, and
one-loop EWK amplitudes.  We have benchmarked them at the level of
amplitude squared against \texttt{OpenLoops
  2}~\cite{Buccioni:2019sur}. We found at least a 12-digit
agreement on the whole grid in all partonic channels.
Second, at the mixed QCD-EWK order, we have checked that all the UV
and IR poles disappear from the finite remainders
$\oF_{i,c}^{(1,1),\rm fin}$ defined in sec~\ref{sec:uvir}.  With our
high-precision numerical evaluation of the Feynman integrals described
in sec~\ref{sec:comp}, we found $\ep$-poles cancellation to $\sim$50
digits on the whole grid in all partonic channels.
Finally, to validate our computational framework beyond the universal
UV and IR structure, we have applied it to the computation of two-loop
massless QCD corrections and compared our findings against results
available in the literature~\cite{Gehrmann:2022vuk,Garland:2002ak}.
For this check, we picked a representative point in the
$u\bar u\to g Z$ channel and targeted a
precision for the finite remainders of about 16 digits. We found
perfect agreement.\footnote{We note that the QCD-QED part of the full
QCD-EWK correction cannot be checked against a trivial abelianisation
of $\mathcal O(\alpha_s^2)$ corrections. Because of this, the NNLO QCD
check required a dedicated calculation. Still, we have modified our
framework as little as possible.}

All the tree-level, one-loop QCD, one-loop EWK, and two-loop mixed
QCD-EWK finite remainders $\oF_{i,c}^{(i,j),\rm fin}$ evaluated on the
grid~\ref{eq:grid} with the renormalisation scale set to
\begin{equation}
  \mu_R = \sqrt{p_{T,Z}^2+m_{Z}^2}+ |p_{T,Z}| \,,
  \label{eq:scale}
\end{equation}
can be found in a ancillary files. For illustrative purposes, here we
show results for the helicity amplitudes in the $u\bar u\to gZ$ channel
at a typical point. In particular, we choose the grid point number
805, i.e.
\begin{equation}
  \{s_{23},s_{13}\} = \left\{-\frac{1327912559351427052473827}{3254131111768750000},-\frac{1647987026840591434577097}{325413111176875000}\right\}.
\end{equation}
We also set the azimuthal angle $\phi$ of eq.~\ref{eq:mompar} to
$\pi/4$.  We parametrise the leptons 5 and 6 in the $Z$ rest frame,
with polar angle $\theta_l$ such that $\cos\theta_l=-0.754$ and
azimuthal angle $\phi_l = \pi/3$. The full kinematics then reads
\begin{equation}
  \begin{split}
    &p_1 = \{1170.54,0,0,+1170.54\},\\
    &p_2 = \{1170.54,0,0,-1170.54\},\\
    &p_{3} = -p_{3,\rm phys} = \{-1168.77,-434.205,-434.205,994.458\},\\
    &p_{5} = -p_{5,\rm phys} = \{-766.262,264.02,253.058,-673.36\},\\
    &p_{6} = -p_{6,\rm phys} = \{-406.056,170.184,181.147,-321.099\}.
  \end{split}
\end{equation}
To present the helicity amplitudes, we define the finite part of
eq.~\ref{eq:hels} as
\begin{equation}
  \begin{split}
  \mathcal M^{\rm fin}_{\vec\lambda} =&
  T^{a_3}_{i_2 i_1} \sqrt{\frac{\alpha_s}{2\pi}}\sqrt{\frac{\alpha}{\pi}}
  \times {\rm pref}_{\vec\lambda} \times\\
  &\left[
    1 + \frac{\alpha}{2\pi} \mathcal M_{\vec\lambda}^{(0,1), \rm fin}
    +\frac{\alpha_s}{2\pi} \mathcal M_{\vec\lambda}^{(1,0), \rm fin}
    +\frac{\alpha}{2\pi}\frac{\alpha_s}{2\pi}
    \mathcal M_{\vec\lambda}^{(1,1), \rm fin} +
    \mathcal O(\alpha^2,\alpha_s^2)
    \right],
  \end{split}
  \label{eq:calMfin}
\end{equation}
where $\alpha_s=\alpha_s(\mu)$ and $\mathcal M^{(i,j), \rm fin}$ are
computed from eq.~\ref{eq:hels} using the corresponding
$\oF^{(i,j),\rm fin}_{i,c}$ tensors defined in sec.~\ref{sec:uvir}. We
re-absord the tree-level amplitude in the ``pref$_{\vec\lambda}$''
terms, which are given in tab.~\ref{tab:pref} for half of the
helicities. Results for the $\{\lambda_1,\lambda_2,\lambda_3,-,+\}$
helicities are obtained from the ones for
$\{\lambda_1,\lambda_2,\lambda_3,+,-\}$ by exchanging
$5\leftrightarrow 6$ and replacing $c_{l,L}\to c_{l,R}$. We report the
results for all the amplitudes in eq.~\ref{eq:calMfin} in
tab.~\ref{tab:calMfin}, for the scale choice eq.~\ref{eq:scale}.

\begin{table}[ht]
  \begin{center}
  \begin{tabular}{c|c}
    & pref$_{\vec\lambda}$\\\hline\hline
    $-+-+-$ & $-16\pi^2\, c_{l,L}\,g_{u,L}
    \langle25\rangle^2[65]/\langle13\rangle/\langle23\rangle$ \\\hline
    $-+++-$ & $- 16\pi^2\, c_{l,L}\,g_{u,L}
    \langle56\rangle[61]^2/[31]/[32]$\\\hline
    $+--+-$ & $16\pi^2\, c_{l,L}\,g_{u,R}
    \langle15\rangle^2[65]/\langle13\rangle/\langle23\rangle$\\\hline
    $+-++-$ & $16\pi^2\, c_{l,L}\,g_{u,R}
    \langle56\rangle[62]^2/[31]/[32]$\\\hline
  \end{tabular}
  \end{center}
  \caption{Tree-level prefactors in the $u\bar u$ channel, see
    eq.~\ref{eq:calMfin} and text for details.}
  \label{tab:pref}
  \end{table}

\begin{table}[ht]
  \begin{center}
  \begin{tabular}{c|c|c|c}
    &  $\mathcal M^{(0,1),\rm fin}$ & $\mathcal M^{(1,0),\rm fin}$ & $\mathcal M^{(1,1),\rm fin}$
    \\
    \hline\hline
    $-+-+-$ & -85.0997-24.7235 i & 0.100766 +0.0657261 i & 126.854 +14.0937 i\\\hline
    $-+++-$ & -42.6517-31.4408 i & -9.71342+5.28643 i & 1155.14 -266.036 i\\\hline
    $+--+-$ & 2.95756 -1.7408 i & 0.784688 -0.579024 i & 29.6715 +1.48943 i\\\hline
    $+-++-$ & -0.886388-0.430958 i & 1.1077 +1.90476 i & 3.82559 -18.7726 i\\\hline
    $-+--+$ & -92.4671-26.3319 i & 0.844284 +1.3139 i & 99.1731 -76.5044 i\\\hline
    $-++-+$ & -72.2596-25.7339 i & 0.981344 +4.53538 i & 244.584 -403.143 i\\\hline
    $+---+$ & 4.0673 -1.4704 i & -9.42891-4.09849 i & 43.8812 +22.9738 i\\\hline
    $+-+-+$ & -1.47287-0.775214 i & -0.0801866+0.592326 i & 2.3056 -13.4416 i
  \end{tabular}
  \end{center}
  \caption{Finite amplitudes in the $u\bar u$ channel, see
    eq.~\ref{eq:calMfin} and text for details.}
  \label{tab:calMfin}
\end{table}

\begin{figure}[h]
		\centering
		\begin{subfigure}[b]{0.45\textwidth}
			\includegraphics[width=\textwidth]{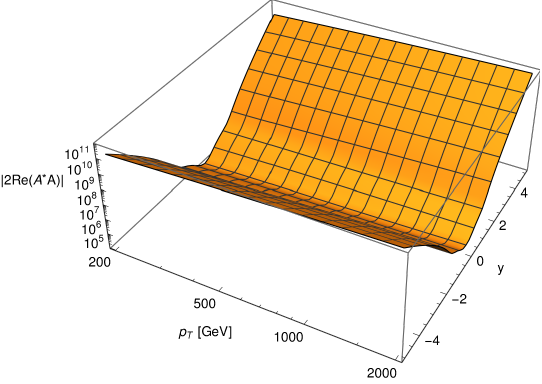}
			\caption{$u \, \bar{u} \to g \, Z$}
		\end{subfigure}
		\begin{subfigure}[b]{0.45\textwidth}
			\includegraphics[width=\textwidth]{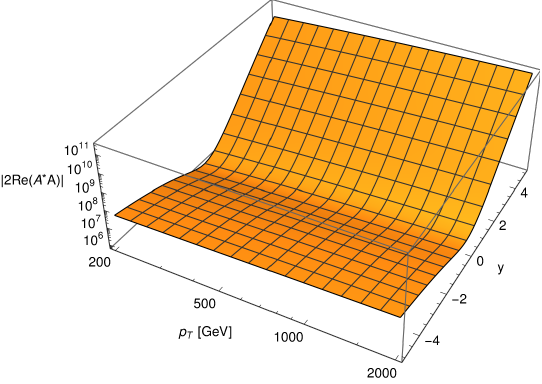}
			\caption{$u \, g \to u \, Z$}
		\end{subfigure}
		\begin{subfigure}[b]{0.45\textwidth}
			\includegraphics[width=\textwidth]{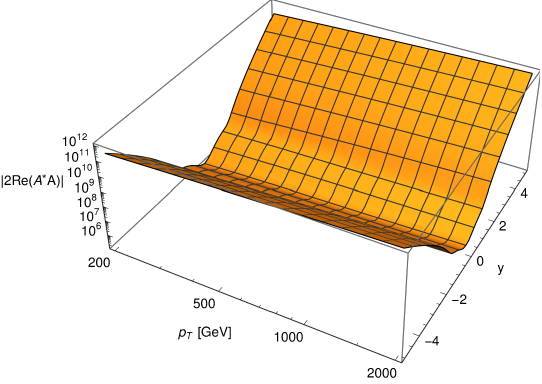}
			\caption{$d \, \bar{d} \to g \, Z$}
		\end{subfigure}
		\begin{subfigure}[b]{0.45\textwidth}
			\includegraphics[width=\textwidth]{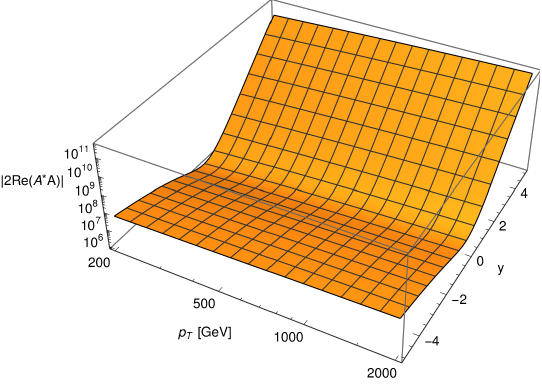}
			\caption{$d \, g \to d \, Z$}
		\end{subfigure}
		\caption{Absolute value of the virtual NNLO finite
                  remainders summed over colour and polarisations, see
                  eq.~\ref{eq:fig}, in all partonic channels as
                  functions of transverse momentum $p_t$ and rapidity
                  $y$ of the $Z$ boson. For simplicity, in this plot we set
                  $\alpha_s=0.118$.}
		\label{fig:num}
\end{figure}

Finally, to illustrate the behaviour of the two-loop mixed QCD-EWK amplitudes
across the whole kinematic coverage that we consider, in fig.~\ref{fig:num} we
plot
\begin{equation}
  \left|\sum_{\rm col}\sum_{\rm pol}2\Re \left[\mathcal A^{(0,0)*} \mathcal
    A^{(1,1),\rm fin}\right]\right|,
  \label{eq:fig}
\end{equation}
see eq.~\ref{eq:ampdef_ren}, as a function of the $Z$-boson transverse
momentum and rapidity, for all the independent partonic channels. The
$(i,j)$ superscript in $\mathcal A^{(i,j)}$ indicates that only the
$A^{(i,j)}$ contribution to eq.~\ref{eq:ampdef_ren} is kept, see
eq.~\ref{eq:ArenDefs}. As it was the case for $\mathcal M^{\rm fin}$,
the ``fin'' superscript implies that we only consider the finite part of
the amplitude, as defined in sec.~\ref{sec:uvir}.

\section{Conclusions and outlook}
\label{sec:concl}

In this paper, we have performed a first important step towards the
calculation of mixed QCD-EWK corrections for boosted $Z$ production in
association with one hard jet. In particular, we have computed for the
first time the bosonic contributions to the two-loop mixed QCD-EWK
amplitudes. Our calculation relies on a recently proposed tensor
decomposition method which reduces the redundancy stemming from
unphysical dimensional regularization remnants. Moreover, we exploited
the modern \texttt{AMFlow} method~\cite{Liu:2022chg} for highly
efficient numerical evaluation of Feynman integrals. We have
numerically evaluated the finite part of the two-loop amplitudes on a
two-dimensional grid in $\{p_{t,Z},y_Z\}$ designed to offer good
coverage for phenomenological investigations at the LHC and future
colliders.  We have performed extensive checks on our
calculation, and expect the precision our final results to be more
than sufficient for phenomenological applications.  Our numerical
evaluation of the tree-level, one-, and two-loop finite remainders in
all the relevant partonic channels are provided in ancillary files.

Our framework is easily generalisable, and we expect it to be able
to cope with the case of fermionic contribution as well, which we
plan to investigate in the future. This would complete the calculation
of the mixed QCD-EWK two-loop amplitudes for this process and will open
the way for interesting investigations. In particular, one could
apply these results to phenomenological studies at the LHC. This 
requires devising appropriate IR subtraction schemes for mixed QCD-QED
real-emission, e.g. along the lines of ref.~\cite{Buccioni:2022kgy}.
A successful completion of this programme would lead to a significant
decrease on the theoretical uncertainty for important LHC analysis, like
boosted Drell-Yan studies or monojet searches. It would also allow for
a thorough study of the onset of the Sudakov regime, where EWK corrections
are dominated by large logarithms. Insight in the transition region can
provide important clues on the structure and size of subleading terms,
and inform Sudakov-based approximations to mixed QCD-EWK corrections for
more complicated processes. We leave these interesting avenues of
exploration to the future.

\section*{Acknowledgements}
We are grateful to L. Tancredi for providing benchmark results for the
tensor decomposition and the form factors for the two-loop QCD result.
We also thank F. Buccioni for providing all the relevant
renormalisation factors in computer-readable form, as well as for
several discussions and comments on the manuscript.  We acknowledge
interesting discussions with A. Penin and J. Lindert on the structure
of the EWK Sudakov logarithms.  The research of PB, FC, HC, and XL was
supported by the ERC Starting Grant 804394 \textsc{HipQCD} and by the
UK Science and Technology Facilities Council (STFC) under grant
ST/T000864/1.  The work of PB was also supported by the Swiss National
Science Foundation (SNF) under contract 200020-204200 and by the
European Research Council (ERC) under the European Union's Horizon
2020 research and innovation programme grant agreement 101019620 (ERC
Advanced Grant TOPUP). HC was also supported in part by
the U.S. Department of Energy under grant DE-SC0010102.
Feynman graphs were drawn with
\texttt{qgraf-xml-drawer}~\cite{qraf:drawer}.

\appendix

\section{Projector coefficients}
\label{app:proj}
For convenience, we provide here an explicit form of the coefficients
$c_{ik}$ required to define in eq.~\ref{eq:proj} the projectors
$\mathcal P_i$ onto our Lorentz tensor basis $\oT_k$.
\begin{equation}
  \label{eq:app.projs}
\begin{split}
  c_{ik} &
= \frac{1}{(d-3) \, tu} \times \\
&
\left(
\renewcommand{\arraystretch}{1.5}
\begin{array}{ccccccc}
\frac{(s+u)^2}{2 s^2} & -\frac{(s+u)^2}{2 s^2 t} & \frac{m_Z^2 s-t u}{2 s^2} & 0 & \frac{t u-m_Z^2 s}{2 s^2 t} & 0 & 0 \\
-\frac{(s+u)^2}{2 s^2 t} & \frac{d (s+u)^2}{2 s^2 t^2} & \frac{t u-m_Z^2 s}{2 s^2 t} & -\frac{u (s+u)}{2 s^2 t} & \frac{d (m_Z^2 s-t u)+2 t (u-s)}{2 s^2 t^2} & \frac{s+u}{2 s t} & 0 \\
\frac{m_Z^2 s-t u}{2 s^2} & \frac{t u-m_Z^2 s}{2 s^2 t} & \frac{(s+t)^2}{2 s^2} & 0 & -\frac{(s+t)^2}{2 s^2 t} & 0 & 0 \\
0 & -\frac{u (s+u)}{2 s^2 t} & 0 & \frac{u^2}{2 s^2} & -\frac{u (s+t)}{2 s^2 t} & 0 & 0 \\
\frac{t u-m_Z^2 s}{2 s^2 t} & \frac{d (m_Z^2 s-t u)+2 t (u-s)}{2 s^2 t^2} & -\frac{(s+t)^2}{2 s^2 t} & -\frac{u (s+t)}{2 s^2 t} & \frac{d (s+t)^2-4 s t}{2 s^2 t^2} & \frac{s-t}{2 s t} & 0 \\
0 & \frac{s+u}{2 s t} & 0 & 0 & \frac{s-t}{2 s t} & \frac{1}{2} & 0 \\
0 & 0 & 0 & 0 & 0 & 0 & \frac{1}{2 (d-4)} \\
\end{array}
\right),
\end{split}
\end{equation}
with $s = s_{12}$, $t = s_{13}$,
$u = s_{23}$.

\section{Integral topologies}
\label{app:topo}
In this appendix we report the definition of the 18 basics
topologies introduced in sec.~\ref{sec:comp}. They read

\scalebox{0.65}{\parbox{1.0\linewidth}{
    \begin{equation}
      \begin{split}
	\text{PL2A}&=\{k_1^2,k_2^2,(k_1-k_2)^2,(k_1-p_1)^2,(k_2-p_1)^2,(k_1-p_{12})^2,(k_2-p_{12})^2,(k_1-p_{123})^2,(k_2-p_{123})^2\}\,, \\ \text{PL2Z1}&=\{k_1^2-m_Z^2,k_2^2,(k_1-k_2)^2,(k_1-p_1)^2,(k_2-p_1)^2,(k_1-p_{12})^2,(k_2-p_{12})^2,(k_1-p_{123})^2,(k_2-p_{123})^2\}\,, \\ \text{PL2Z3}&=\{k_1^2,k_2^2,(k_1-k_2)^2-m_Z^2,(k_1-p_1)^2,(k_2-p_1)^2,(k_1-p_{12})^2,(k_2-p_{12})^2,(k_1-p_{123})^2,(k_2-p_{123})^2\}\,, \\ \text{PL2Z4}&=\{k_1^2,k_2^2,(k_1-k_2)^2,-m_Z^2+(k_1-p_1)^2,(k_2-p_1)^2,(k_1-p_{12})^2,(k_2-p_{12})^2,(k_1-p_{123})^2,(k_2-p_{123})^2\}\,, \\ \text{PL2W1}&=\{k_1^2-m_W^2,k_2^2,(k_1-k_2)^2,(k_1-p_1)^2,(k_2-p_1)^2,(k_1-p_{12})^2,(k_2-p_{12})^2,(k_1-p_{123})^2,(k_2-p_{123})^2\}\,, \\ \text{PL2W3}&=\{k_1^2,k_2^2,(k_1-k_2)^2-m_W^2,(k_1-p_1)^2,(k_2-p_1)^2,(k_1-p_{12})^2,(k_2-p_{12})^2,(k_1-p_{123})^2,(k_2-p_{123})^2\}\,, \\ \text{PL2W4}&=\{k_1^2,k_2^2,(k_1-k_2)^2,-m_W^2+(k_1-p_1)^2,(k_2-p_1)^2,(k_1-p_{12})^2,(k_2-p_{12})^2,(k_1-p_{123})^2,(k_2-p_{123})^2\}\,, \\ \text{PL2W29}&=\{k_1^2,k_2^2-m_W^2,(k_1-k_2)^2,(k_1-p_1)^2,(k_2-p_1)^2,(k_1-p_{12})^2,(k_2-p_{12})^2,(k_1-p_{123})^2,-m_W^2+(k_2-p_{123})^2\}\,, \\ \text{NPL2A}&=\{k_1^2,k_2^2,(k_1-k_2)^2,(k_1-p_1)^2,(k_2-p_1)^2,(k_1-p_{12})^2,(k_1-k_2+p_3)^2,(k_2-p_{123})^2,(k_1-k_2-p_{12})^2\}\,, \\ \text{NPL2Z1}&=\{k_1^2-m_Z^2,k_2^2,(k_1-k_2)^2,(k_1-p_1)^2,(k_2-p_1)^2,(k_1-p_{12})^2,(k_1-k_2+p_3)^2,(k_2-p_{123})^2,(k_1-k_2-p_{12})^2\}\,, \\ \text{NPL2Z4}&=\{k_1^2,k_2^2,(k_1-k_2)^2,-m_Z^2+(k_1-p_1)^2,(k_2-p_1)^2,(k_1-p_{12})^2,(k_1-k_2+p_3)^2,(k_2-p_{123})^2,(k_1-k_2-p_{12})^2\}\,, \\ \text{NPL2Z7}&=\{k_1^2,k_2^2,(k_1-k_2)^2,(k_1-p_1)^2,(k_2-p_1)^2,(k_1-p_{12})^2,-m_Z^2+(k_1-k_2+p_3)^2,(k_2-p_{123})^2,(k_1-k_2-p_{12})^2\}\,, \\ \text{NPL2Z1c13c24}&=\{k_1^2-m_Z^2,k_2^2,(k_1-k_2)^2,(k_1-p_3)^2,(k_2-p_3)^2,(k_1+p_1+p_2)^2,(k_1-k_2+p_1)^2,(k_2+p_2)^2,(k_1-k_2+p_1+p_2)^2\}\,, \\ \text{NPL2W1}&=\{k_1^2-m_W^2,k_2^2,(k_1-k_2)^2,(k_1-p_1)^2,(k_2-p_1)^2,(k_1-p_{12})^2,(k_1-k_2+p_3)^2,(k_2-p_{123})^2,(k_1-k_2-p_{12})^2\}\,, \\ \text{NPL2W4}&=\{k_1^2,k_2^2,(k_1-k_2)^2,-m_W^2+(k_1-p_1)^2,(k_2-p_1)^2,(k_1-p_{12})^2,(k_1-k_2+p_3)^2,(k_2-p_{123})^2,(k_1-k_2-p_{12})^2\}\,, \\ \text{NPL2W7}&=\{k_1^2,k_2^2,(k_1-k_2)^2,(k_1-p_1)^2,(k_2-p_1)^2,(k_1-p_{12})^2,-m_W^2+(k_1-k_2+p_3)^2,(k_2-p_{123})^2,(k_1-k_2-p_{12})^2\}\,, \\ \text{NPL2W1c13c24}&=\{k_1^2-m_W^2,k_2^2,(k_1-k_2)^2,(k_1-p_3)^2,(k_2-p_3)^2,(k_1+p_1+p_2)^2,(k_1-k_2+p_1)^2,(k_2+p_2)^2,(k_1-k_2+p_1+p_2)^2\}\,, \\ \text{NPL2W28}&=\{k_1^2,k_2^2-m_W^2,(k_1-k_2)^2,(k_1-p_1)^2,(k_2-p_1)^2,(k_1-p_{12})^2,(k_1-k_2+p_3)^2,-m_W^2+(k_2-p_{123})^2,(k_1-k_2-p_{12})^2\},
      \end{split}
      \nonumber
    \end{equation}
    \label{eq:qqgZ.topos}
}}

\noindent
where each list corresponds to $\{\mathcal D_{t,1},...,\mathcal D_{t,9}\}$,
see sec.~\ref{sec:comp}.
For convenience, these topologies definitions are also provided in
computer-readable format in an ancillary file. 

\bibliographystyle{JHEP}
\bibliography{ppjZ}

\end{document}